\documentclass[cameraready]{Interspeech}

\title{Multilingual Multi-Speaker Unit Vocoders: A Systematic Analysis of  \\ Discrete Speech Representations
}

\author[affiliation={1}]{Naman}{Kothari}
\author[affiliation={2}]{Arjun}{Gangwar}
\author[affiliation={2}]{Adarsh}{Arigala}
\author[affiliation={2}]{S}{Umesh}

\address{
    $^1$ National Institute of Technology, Trichy \\
    $^2$ Indian Institute of Technology, Madras
}

\email{namank.edu@gmail.com, arjungangwar@gmail.com, arigalaadarsh780@gmail.com, umeshs@ee.iitm.ac.in}


\usepackage{comment}
 
\usepackage{tabularx}             
\usepackage{multirow}             
\usepackage{graphicx}             
\usepackage{caption}              

\usepackage{pdflscape}            
\newcolumntype{Y}{>{\centering\arraybackslash}X}  
\usepackage{float} 
\usepackage{makecell}
 
\begin{document}

\maketitle




\begin{abstract}

Discrete speech units obtained via k-means clustering of self supervised embeddings entangle phonetic, speaker, and language information, causing speaker mixing and cross-lingual interference in multilingual multi-speaker speech generation. Despite growing use in Audio LLMs and speech to speech systems, unit vocoders remain underexplored. We analyze a BigVGAN based unit vocoder, across four Indian languages. We study the interaction between cluster size and conditioning strategies using WER, speaker similarity, and unit level metrics. Results show that cluster size governs intelligibility by improving phonetic discriminability, while explicit speaker conditioning is indispensable for preventing identity collapse. Language supervision yields further gains mainly at lower cluster sizes where units remain ambiguous. Our analysis shows similar phonemes across languages collapse to the same cluster IDs at smaller inventories, with larger clusters progressively separating them.

\end{abstract}
\footnote{ We will be releasing the full codebase and trained models.}

\begin{figure*}[!t]  
    \centering
    \includegraphics[width=\linewidth]{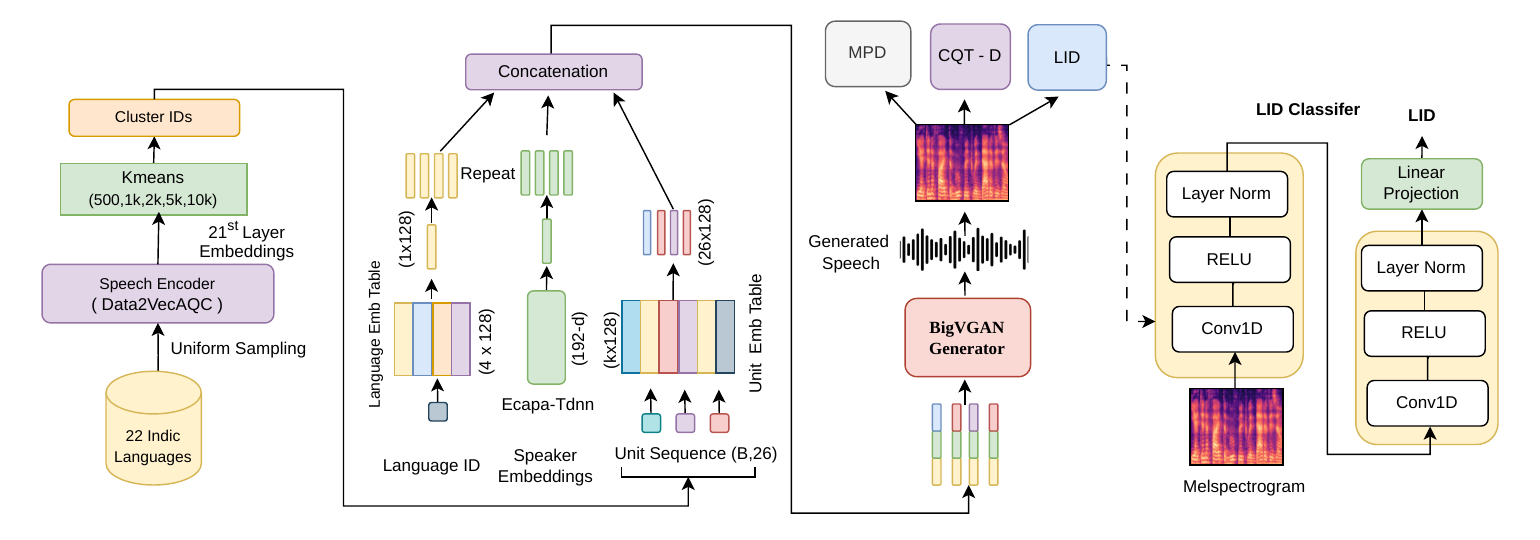}
    \captionsetup{font=small}
    \caption{Overview of the proposed method.}
    \label{fig:architecture} 
  
\end{figure*}

\section{Introduction}

Discrete speech units derived from self-supervised representations are emerging as a key intermediate in modern speech systems, particularly in speech language models and speech-to-speech translation pipelines. By discretizing continuous speech into symbolic unit sequences, these approaches enable direct speech generation without relying on intermediate text, making them attractive for multilingual and unwritten language settings. However, while research focuses on the upstream task of mapping speech to discrete units, the downstream unit vocoder responsible for waveform reconstruction is often treated as a secondary component. Consequently, the interaction between unit sharing, conditioning, and cluster size remains underexplored in multilingual multi-speaker settings.

Early work on unit based resynthesis~\cite{DBLP:journals/corr/abs-2104-00355} investigates reconstructing speech from discrete acoustic representations. While this work demonstrates the feasibility of high quality resynthesis, it is restricted to English. It also highlights limitations of fixed speaker embedding lookup tables and instead adopts d vectors, and leaves multilingual behavior and cross language consistency unexplored. Recent multilingual S2ST systems~\cite{gong2023multilingualspeechtospeechtranslationmultiple} adopt multi-speaker multilingual setups but rely on separate SSL and clustering models for different language families, limiting cross-lingual consistency. Speaker conditioning is typically implemented through embedding lookup tables tied to training speaker identities, and evaluation focuses mainly on WER, with limited analysis of perceptual quality, speaker similarity, or larger cluster sizes. Large scale systems such as Seamless~\cite{communication2023seamlessmultilingualexpressivestreaming} extend speech translation to many languages and speakers, but evaluation primarily emphasizes translation performance with limited analysis of vocoder specific aspects such as perceptual quality and speaker consistency. Similarly, SpeechGPT~\cite{zhang-etal-2023-speechgpt} supports multi speaker modeling but remains monolingual and lacks systematic analysis of cross lingual unit behavior. Streaming approaches such as StreamSpeech~\cite{zhang-etal-2024-streamspeech} focus on monolingual single speaker settings, using collapsed discrete units that require a duration predictor within the vocoder and introduce potential alignment errors.

Despite the widespread use of discrete units in speech synthesis and multilingual translation, unit vocoders are rarely studied independently of larger system objectives. Systematic analysis of conditioning mechanisms and cluster size choices in multilingual multi-speaker settings remains limited, particularly at larger cluster inventories. Moreover, most studies focus on English and rely on HiFi-GAN \cite{NEURIPS2020_c5d73680} based architectures rather than newer vocoder designs.

In this work, we study discrete unit vocoders as the primary object of analysis in a multilingual multi-speaker setting. We focus on four Indian languages (Bengali, Hindi, Tamil, and Telugu) from two linguistically distinct families, Indo-Aryan and Dravidian, and examine how clustering choices and conditioning strategies influence intelligibility, speaker consistency, and phoneme-level structure. \textbf{Our contributions}: 
(i) We extend the BigVGAN architecture \cite{lee2023bigvgan} to a multilingual multi-speaker discrete unit vocoder framework.
(ii) We investigate the role of speaker and language conditioning and their impact on intelligibility and speaker preservation.
(iii) Within this setup, we systematically analyze cluster size trade offs and cross-lingual unit sharing in a unified multilingual setting.
(iv) We provide a comprehensive evaluation beyond WER using speaker similarity and unit level metrics to better characterize the learned units.

Our analysis provides practical insights into configuration choices for multilingual multi-speaker unit vocoders and clarifies how conditioning mechanisms and cluster size influence speech generation quality.

\vspace{-0.5cm}
\section{Methodology}
For our experiments, we adopt the BigVGAN \cite{lee2023bigvgan} architecture consisting of a generator $\mathcal{G}$, a Multi-Period Discriminator (MPD) and CQT based time frequency discriminators. Unlike a standard vocoder, which takes mel spectrograms as input, we replace the generator input with discrete units. Optional speaker and language embeddings are concatenated to the unit representations before being passed to the generator as shown in Fig \ref{fig:architecture}.

\subsection{Generator Conditioning}
During training, 16 kHz audio is split into fixed length segments of 8320 samples. Frame level units are extracted with a hop size of 320 samples, yielding 26 units per segment, where cluster IDs lie in $\{0, \ldots, k-1\}$, for a cluster size $k$, with an additional index $k$ used for padding. Hence, the generator takes unit sequences of shape $[B, 26]$, where $B$ is the batch size. During inference the generator operates on full-length unit sequences.

Let $\mathbf{c} \in \mathbb{N}^{B \times T}$ with $T = 26$ denote a batch of discrete units. 
Units are mapped through a learned embedding table as 
$\mathbf{U} = \mathbf{E}_u[\mathbf{c}] \in \mathbb{R}^{B \times d_u \times T}$, 
where $d_u = 128$.

For speaker conditioning, we obtain a speaker embedding 
$\mathbf{S} \in \mathbb{R}^{B \times d_s}$ with $d_s = 192$ 
using a pre-trained ECAPA-TDNN \cite{Dawalatabad_2021}. Unlike fixed speaker lookup tables, which are restricted to the training set, ECAPA-TDNN provides a continuous speaker space that improves the model's generalization to unseen speakers and better preserves identity in a multilingual context.

For language conditioning, given language IDs 
$\ell \in \{0, \dots, N-1\}$ for each sample, we obtain embeddings via a learnable lookup table as 
$\mathbf{L} = \mathbf{E}_{\text{lang}}[\ell] \in \mathbb{R}^{B \times d_\ell}$, 
where $d_\ell = 128$. Both embeddings are repeated along the temporal dimension to match the unit sequence length, yielding 
$\tilde{\mathbf{S}} \in \mathbb{R}^{B \times d_s \times T}$ and 
$\tilde{\mathbf{L}} \in \mathbb{R}^{B \times d_\ell \times T}$.
The generator input is formed via channel-wise concatenation:
$
\mathbf{X} = \text{Concat}(\tilde{\mathbf{L}}, \tilde{\mathbf{S}}, \mathbf{U})
 $
where speaker or language terms are omitted when not used. The waveform is synthesized as $ 
\hat{\mathbf{y}} = \mathcal{G}(\mathbf{X})
$
 
\subsection{Auxiliary LID Classification}
A common issue in multilingual modeling is language interference. To mitigate this, we employ a Language Identification (LID) Classifier similar to \cite{gong2023multilingualspeechtospeechtranslationmultiple} which operates on mel spectrograms instead of raw waveforms, to predict the language ID. The LID classifier consists of two convolutional layers, each followed by ReLU activation and LayerNorm. Finally, a linear projection is applied on top of the second convolutional layer to predict the language label.

When enabled, the LID classifier is trained jointly with the vocoder using a separate optimizer. It is supervised on real mel-spectrograms using a cross-entropy loss. The same loss is also computed on mel-spectrograms generated by the vocoder and incorporated into the generator objective, weighted by $\lambda_{\text{LID}}$. This encourages the generator to produce speech that is consistent with the target language.

\subsection{Training Objective}
 
We follow the adversarial training formulation of BigVGAN. Both discriminators are trained using the least squares GAN objective \cite{Mao_2017_ICCV}. In addition to the adversarial loss, we use feature matching loss \cite{pmlr-v48-larsen16} and an $\ell_1$ mel spectrogram reconstruction loss, as in BigVGAN.
\noindent The overall generator objective is given  
\setlength{\abovedisplayskip}{4pt}
\setlength{\belowdisplayskip}{4pt}
\[\mathcal{L}_{\mathcal{G}}
=
\mathcal{L}_{\text{adv}}
+
\lambda_{\text{fm}}\mathcal{L}_{\text{fm}}
+
\lambda_{\text{mel}}\mathcal{L}_{\text{mel}}
+
\lambda_{\text{LID}}\mathcal{L}_{\text{LID}}^{\text{gen}},
\] where $\mathcal{L}_{\text{adv}}$, $\mathcal{L}_{\text{fm}}$, and $\mathcal{L}_{\text{mel}}$ follow the definitions in BigVGAN, and $\lambda_{\text{fm}}$ and $\lambda_{\text{mel}}$ are set to 1 and 15, respectively.

Let $f_{\operatorname{LID}}(\cdot)$ denote a language classifier.
The classifier is trained on real mel spectrograms using cross-entropy:
$
\mathcal{L}_{\text{LID}}^{\text{real}}
=
\operatorname{CE}(\hat{\mathbf{y}}_{\text{real}}, \mathbf{y})
$.
During generator updates, the same loss is computed on generated samples:
$
\mathcal{L}_{\text{LID}}^{\text{gen}}
=
\operatorname{CE}(\hat{\mathbf{y}}_{\text{gen}}, \mathbf{y})
$
where $\hat{\mathbf{y}}_{\text{real}}$ and $\hat{\mathbf{y}}_{\text{gen}}$ denote the logits from the LID classifier when real and vocoder-generated mel spectrograms are passed through it, respectively, and $\mathbf{y}$ denotes the ground truth language labels. This auxiliary term $\mathcal{L}_{\text{LID}}^{\text{gen}}$ is added to the generator objective to encourage language-consistent synthesis. In our experiments, we set $\lambda_{\text{LID}} = 1$.




\subsection{Multilingual Unit Extraction}

We use Data2Vec-AQC \cite{lodagala2023data2vecaqcsearchrightteaching}, an SSL model pretrained on 30,000 hours of speech in 23 Indian languages, with strong coverage of Bengali, Hindi, Tamil, and Telugu. This model achieves the best reported performance on several Indian ASR benchmarks when fine tuned, making it well suited for learning robust multilingual speech representations \cite{lodagala2024all}. We use 21st layer representations for unit extraction, as deeper SSL layers capture higher level semantic information \cite{pasad2023comparative}. This is particularly well-suited for speech generation within speech-to-speech translation and Audio LLM pipelines, where semantic units serve as the primary intermediate representation.

We train five multilingual k-means models with cluster sizes of 500, 1k, 2k, 5k, and 10k using representations extracted from the 21st layer of Data2Vec-AQC. 
The clustering is conducted on 1,200 hours of speech across 22 Indian languages, with \textbf{balanced} sampling across languages. The data is drawn from publicly available corpora, including IndicVoices, IndicTTS, Shrutilipi, and SPRING-INX~\cite{javed-etal-2024-indicvoices,kumar2023buildingtexttospeechsystemsbillion,DBLP:conf/icassp/BhogaleRJDKKK23,r2023springinxmultilingualindianlanguage}.

\section{Experiments}

\subsection{Training Setup and Datasets}

We conduct experiments on four languages Bengali, Hindi, Tamil, and Telugu from the IndicVoices-R dataset \cite{ai4bharat2024indicvoicesr}, a multilingual and multispeaker corpus. All audios are resampled to 16 kHz. Language-wise data and speaker statistics are summarized in Table~\ref{tab:dataset_stats}.  We evaluate on the official IndicVoices-R test splits, comprising 16 unseen speakers per language, ensuring evaluation under speaker-generalization conditions. All models are trained for 400k steps with a batch size of 64 across four NVIDIA A100 GPUs, requiring approximately four days of training. The generator, discriminators, and LID classifier are optimised using independent AdamW optimizers with a learning rate of 0.0001. 

For our experiments, we use the discrete units under four conditioning strategies: units only (unconditioned), adding speaker conditioning using ECAPA TDNN embeddings, adding language conditioning with an auxiliary LID objective, and a combined setting with both speaker and language conditioning.

\begin{table}[!h]
\vspace{-0.2cm}
\centering
\small
\caption{IndicVoices-R training set statistics.}
\label{tab:dataset_stats}
\begin{tabular}{lcc}
\hline
\textbf{Language} & \textbf{Hours} & \textbf{\# Speakers} \\
\hline
Bengali & 109.44 & 585 \\
Hindi   & 71.8   & 368 \\
Tamil   & 97.3   & 1052 \\
Telugu  & 133.9  & 649 \\
\hline
\end{tabular}
\label{tab:lang_stats}
\vspace{-0.5cm}
\end{table}
\subsection{Evaluation Metrics}

We evaluate intelligibility, speaker consistency, and perceptual audio quality using objective metrics including Word Error Rate (WER) and speaker similarity. For intelligibility, generated speech is transcribed using the Indic-Conformer 600M \footnote{Indic-Conformer: https://github.com/AI4Bharat/IndicConformerASR} ASR model, and WER is computed against the ground truth transcripts. Speaker similarity is measured as the cosine similarity between ECAPA-TDNN \cite{Dawalatabad_2021} embeddings of reference and generated speech. As an upper bound, ground-truth similarity is obtained by averaging cosine similarities between speaker embeddings from multiple utterances of the same speaker. We use Versa~\cite{shi2025versa} to compute speaker similarity scores.

\vspace{-0.1cm}
\subsection{Unit Level Metrics}
To further analyze model behavior, we conduct unit-level analysis following~\cite{9585401}, computing Phoneme Purity, cluster purity, and Phone-Normalized Mutual Information (PNMI) by aligning discrete units with ground truth phoneme labels using forced alignment via IndicMFA \footnote{IndicMFA: https://github.com/AI4Bharat/IndicMFA}.
Phoneme purity and PNMI measure the strength of alignment between discovered units and phoneme labels, with higher values indicating lower phoneme uncertainty given unit assignments and more discriminative phonetic structure. Cluster purity measures how consistently frames of the same phoneme are mapped to a single unit, higher values indicate less phoneme fragmentation, while lower values at larger inventories reflect finer sub phonetic granularity. These metrics help relate model performance to the phonetic structure captured by the learned units.

\section{Results and Discussion}
This section analyzes how cluster size and conditioning strategies affect intelligibility and speaker preservation. Through a systematic comparison of four configurations, we study the individual and combined effects of cluster size, speaker conditioning, and language supervision, relating these trends to underlying unit quality metrics.

\subsection{Impact of Conditioning Strategies}

\textbf{Unit only (Unconditioned):} When the vocoder is trained on multilingual units without explicit speaker or language conditioning, WER decreases steadily with increasing cluster size. For example, Bengali WER falls from 60.42 at 500 clusters to 25.13 at 10k clusters, with similar reductions observed for the other languages (Table \ref{tab:WER_table}-(i)). However, generated audio exhibits strong intra utterance speaker intermixing, often producing audible gender switches. Correspondingly, speaker similarity remains low, typically around 0.16 to 0.21 (Table \ref{tab:spk_sim_table}-(i)), indicating poor preservation of speaker identity.

\textbf{Speaker Conditioning:} Concatenating ECAPA TDNN speaker embeddings to the unit embedding channels produces large gains in speaker similarity, increasing it by approximately 4–5×, depending on language and cluster size (Table~\ref{tab:spk_sim_table}-(ii)), reaching values between 0.67 and 0.77 at 10k clusters. WER also improves across cluster sizes (Table \ref{tab:WER_table}-(ii)),  indicating that explicit speaker conditioning disentangles speaker identity from the unit representations and stabilizes generation.

\begin{table}[!h]
\centering
\caption{WER (\%) across conditioning and clusters (k)}
\label{tab:WER_table}
\resizebox{\columnwidth}{!}{%
\renewcommand{\arraystretch}{1.0} 
\begin{tabular}{l*{7}{>{\centering\arraybackslash}p{0.9cm}}}
\hline
\multirow{2}{*}{\textbf{Language}} & \multicolumn{5}{c}{\textbf{WER  ↓ }} & \multirow{2}{*}{\textbf{\makecell{Ground \\ Truth}}} \\
 
 & \textbf{500} & \textbf{1k} & \textbf{2k} & \textbf{5k} & \textbf{10k} &  \\
\hline
 (i) & \multicolumn{6}{c}{ Units only (Unconditioned)} \\ 
\hline
Bengali & 60.42 & 45.69 & 35.28 & 27.46 & 25.13 & 13.08 \\
Hindi   & 69.46 & 52.96 & 39.51 & 29.47 & 25.31 & 12.57 \\
Tamil   & 86.06 & 80.22 & 71.33 & 62.40 & 59.20 & 30.57 \\
Telugu  & 87.85 & 76.78 & 65.10 & 55.54 & 47.92 & 13.87 \\
\hline
(ii) &\multicolumn{6}{c}{ Units + ECAPA-TDNN speaker conditioning} \\ 
\hline
Bengali & 63.8 & 46.24 & 34.07 & 24.89 & 22.94 & 13.08 \\
Hindi   & 71.52 & 54.14 & 40.47 & 28.17 & 23.99 & 12.57 \\
Tamil   & 86.42 & 77.29 & 65.79 & 56.74 & 51.06 & 30.57 \\
Telugu  & 91.20 & 79.16 & 65.16 & 53.89 & 48.80 & 13.87 \\
\hline
(iii) &\multicolumn{6}{c}{ Units + language embedding + LID loss} \\ 
\hline 
Bengali & 57.49 & 41.97 & 34.21 & 26.33 & 23.94 & 13.08 \\
Hindi   & 64.71 & 49.88 & 36.4 & 28.26 & 26.3 & 12.57 \\
Tamil   & 86.63 & 77.50 & 70 & 63.33 & 58.83 & 30.57 \\
Telugu  & 84.76 & 75.40 & 64.97 & 51.95 & 49.53 & 13.87 \\
\hline
(iv) & \multicolumn{6}{c}{ Units + (ECAPA speaker + language emb) + LID loss} \\ 
\hline 
Bengali & 59.21 & 43.73 & 33.20 & 25.26 & 23.39 & 13.08 \\
Hindi   & 66.57 & 51.34 & 38.28 & 27.63 & 24.84 & 12.57 \\
Tamil   & 84.38 & 76.40 & 68.62 & 54.95 & 52.49 & 30.57 \\
Telugu  & 87.34 & 77.84 & 65.40 & 53.21 & 48.21 & 13.87 \\
\hline 
\end{tabular}
}
 
\vspace{-0.2cm}

\end{table}

\textbf{Language Conditioning:} Incorporating language embeddings with the auxiliary LID objective 
reduces WER at smaller cluster sizes compared to the unconditioned setup, 
with the gains diminishing as cluster size increases (Table~\ref{tab:WER_table}-(iii)). Speaker similarity also improves relative to the unconditioned baseline (Table~\ref{tab:spk_sim_table}-(iii)), but remains notably lower than with explicit speaker conditioning. These results suggest that language conditioning primarily improves linguistic intelligibility when unit granularity is limited, but does not effectively prevent speaker identity collapse.

\textbf{Combining Speaker and Language Conditioning:} Combining speaker embeddings, language embeddings, and the LID objective yields the best overall trade off between intelligibility and speaker similarity. Speaker similarity reaches its highest values, up to 0.78 for Bengali at larger cluster sizes (Table \ref{tab:spk_sim_table}-(iv)). In terms of WER, most of the improvement over the unconditioned model is obtained through speaker conditioning. At smaller cluster sizes such as 1k, adding language supervision on top of speaker embeddings provides additional gains (e.g., Bengali 46.24 with speaker only vs. 43.73 with speaker + language). At larger cluster sizes such as 10k, however,
additional language conditioning yields little benefit and can slightly degrade performance (e.g., Hindi 23.99 vs. 24.84; Table \ref{tab:WER_table}-(iv)).

\begin{table}[!h]
\centering
\caption{Speaker similarity across conditioning and clusters (k)}
\label{tab:spk_sim_table}
\resizebox{\columnwidth}{!}{%
\renewcommand{\arraystretch}{1.0}
\begin{tabular}{l*{6}{>{\centering\arraybackslash}p{0.9cm}}}
\hline
\multirow{2}{*}{\textbf{Language}} & \multicolumn{5}{c}{\hspace{-0.5cm}\textbf{SIM ↑}} & \multirow{2}{*}{\textbf{\makecell{Ground \\ Truth}}}\\
 
 & \textbf{500} & \textbf{1k} & \textbf{2k} & \textbf{5k} & \textbf{10k} & \\
\hline

(i) & \multicolumn{6}{c}{Units only (Unconditioned)} \\
\hline
Bengali & 0.16 & 0.17 & 0.18 & 0.18 & 0.19 & 0.72 \\
Hindi   & 0.16 & 0.19 & 0.18 & 0.19 & 0.18 & 0.63 \\
Tamil   & 0.20 & 0.21 & 0.21 & 0.21 & 0.21 & 0.70 \\
Telugu  & 0.19 & 0.20 & 0.21 & 0.21 & 0.21 & 0.69 \\
\hline

(ii) & \multicolumn{6}{c}{Units + ECAPA-TDNN speaker conditioning} \\
\hline
Bengali & 0.72 & 0.75 & 0.76 & 0.75 & 0.77 & 0.72 \\
Hindi   & 0.67 & 0.70 & 0.72 & 0.70 & 0.73 & 0.63 \\
Tamil   & 0.67 & 0.68 & 0.71 & 0.70 & 0.71 & 0.70 \\
Telugu  & 0.69 & 0.72 & 0.73 & 0.73 & 0.74 & 0.69 \\
\hline

(iii) & \multicolumn{6}{c}{Units + language embedding + LID loss} \\
\hline
Bengali & 0.38 & 0.40 & 0.38 & 0.38 & 0.40 & 0.72 \\
Hindi   & 0.30 & 0.32 & 0.32 & 0.33 & 0.34 & 0.63 \\
Tamil   & 0.36 & 0.37 & 0.35 & 0.37 & 0.36 & 0.70 \\
Telugu  & 0.39 & 0.38 & 0.38 & 0.40 & 0.39 & 0.69 \\
\hline

(iv) & \multicolumn{6}{c}{Units + (ECAPA speaker + language emb) + LID loss} \\
\hline
Bengali & 0.71 & 0.75 & 0.78 & 0.78 & 0.78 & 0.72 \\
Hindi   & 0.64 & 0.68 & 0.71 & 0.73 & 0.71 & 0.63 \\
Tamil   & 0.65 & 0.67 & 0.70 & 0.71 & 0.72 & 0.70 \\
Telugu  & 0.70 & 0.71 & 0.74 & 0.75 & 0.76 & 0.69 \\
\hline
\end{tabular}
}
\label{tab:spk_sim_all_versions}
 
\end{table}

\subsection{Further Analysis}

To better understand these trends, we examine phoneme purity, cluster purity, and PNMI (Table \ref{tab:unit_level_metrics}). Phoneme purity and PNMI increase steadily with cluster size, while cluster purity decreases, indicating a transition from coarse, mixed clusters at small inventories to phoneme aligned units at larger clusters.

\begin{table}[h]
\caption{Unit-level analysis across cluster sizes.}
\label{tab:unit_level_metrics}
\centering
\footnotesize
\setlength{\tabcolsep}{4pt}
\renewcommand{\arraystretch}{1.0}

\begin{tabular}{l l c c c c c}
\hline
\textbf{Metric} & \textbf{Language} & \textbf{500} & \textbf{1k} & \textbf{2k} & \textbf{5k} & \textbf{10k} \\
\hline
\multirow{4}{*}{\makecell{Phoneme \\ Purity}}
 & Bengali & 0.336 & 0.383 & 0.431 & 0.499 & 0.539 \\
 & Hindi   & 0.291 & 0.338 & 0.398 & 0.477 & 0.522 \\
 & Tamil   & 0.303 & 0.336 & 0.372 & 0.434 & 0.474 \\
 & Telugu  & 0.311 & 0.349 & 0.384 & 0.452 & 0.502 \\
\hline
\multirow{4}{*}{\makecell{Cluster \\Purity}}
 & Bengali & 0.123 & 0.087 & 0.070 & 0.056 & 0.049 \\
 & Hindi   & 0.126 & 0.088 & 0.070 & 0.053 & 0.045 \\
 & Tamil   & 0.140 & 0.099 & 0.079 & 0.058 & 0.055 \\
 & Telugu  & 0.152 & 0.114 & 0.090 & 0.060 & 0.049 \\
\hline
\multirow{4}{*}{PNMI}
 & Bengali & 0.22 & 0.30 & 0.37 & 0.46 & 0.51 \\
 & Hindi   & 0.19 & 0.26 & 0.34 & 0.44 & 0.50 \\
 & Tamil   & 0.17 & 0.22 & 0.29 & 0.37 & 0.42 \\
 & Telugu  & 0.17 & 0.23 & 0.29 & 0.38 & 0.45 \\
\hline
\end{tabular}

\end{table}

The steady reduction in WER with increasing cluster size is strongly correlated the increase in phoneme purity and PNMI, showing that intelligibility is largely determined by phonetic resolution. At smaller cluster sizes, lower phoneme purity and PNMI together with higher cluster purity indicate many to many phoneme-unit mappings across languages, creating cross language ambiguity where acoustically similar segments share the same cluster IDs. Tamil and Telugu consistently show higher WER, which aligns with their lower phoneme purity and PNMI values and thus weaker phoneme unit alignment.


At lower cluster sizes such as 1k, this ambiguity makes language supervision particularly effective, reducing cross-language confusions and yielding measurable WER gains. As cluster size increases (e.g., 10k), unit representations become more discriminative and the benefit of language supervision diminishes. In some cases, additional conditioning provides no improvement and may even slightly degrade performance.

Overall, cluster size primarily determines intelligibility through phonetic discriminability, while speaker conditioning plays the key role in preserving voice identity. Language supervision provides additional gains mainly when unit representations are coarse and ambiguous.

\subsection{Analyzing Cross-Lingual Unit Sharing}
Table \ref{tab:unit_sharing_table} illustrates cross-lingual sharing by listing the most frequent cluster IDs for common phonemes across languages. At 500 clusters, shared units for acoustically similar phonemes indicate strong cross-lingual overlap, whereas at 10k clusters, these phonemes segregate into language-specific units. For example, the vowels /a\!:/ (aa) and /i/ map to identical cluster IDs across all four languages at 500 clusters, while /a/ partially splits between Indo-Aryan (Bengali, Hindi) and Dravidian (Tamil, Telugu). In contrast, at 10k clusters these phonemes map to distinct cluster IDs for each language, indicating finer phonetic separation. This sharing at smaller cluster sizes creates phonemic ambiguity, contributing to the higher observed WER. The elevated WER for Tamil and Telugu is consistent with the trends in Table \ref{tab:unit_level_metrics}, reflecting weaker phoneme–unit alignment. For example, in Telugu, cluster ID 425 corresponds to both /l/ and /s/, creating intra-language collisions that introduce ambiguity during synthesis. These observations suggest that larger unit inventories produce more phoneme-aligned representations, which are better suited for systems involving multilingual speech generation such as Audio LLMs and speech-to-speech translation.



 
\begin{table}[!h]
\centering
\caption{ cluster IDs for shared phonemes across languages \\ at 500 and 10k clusters.}
\label{tab:unit_sharing_table}
\footnotesize
\setlength{\tabcolsep}{4pt}
\renewcommand{\arraystretch}{0.99}

\begin{tabular}{cccccc}
\hline
\multirow{2}{*}{\textbf{Cluster Size}} & \multirow{2}{*}{\textbf{Phoneme}} & \multicolumn{4}{c}{\textbf{Cluster ID}} \\
\cline{3-6}
 & & \textbf{Bengali} & \textbf{Hindi} & \textbf{Tamil} & \textbf{Telugu} \\
\hline

\multirow{5}{*}{500}
 & /a/  & 5   & 5   & 179 & 179 \\
 & /a\!:/ & 5   & 5   & 5   & 5 \\
 & /i/  & 281 & 281 & 281 & 281 \\
 & /l/  & 22  & 22  & 287 & 425 \\
 & /s/  & 425 & 425 & 425 & 425 \\
\hline

\multirow{5}{*}{10k}
 & /a/  & 6301 & 4853 & 6596 & 3639 \\
 & /a\!:/ & 1810 & 6013 & 4846 & 8837 \\
 & /i/  & 9586 & 3336 & 8065 & 8065 \\
 & /l/  & 4074 & 2922 & 1350 & 676 \\
 & /s/  & 5144 & 6424 & 8319 & 9441 \\
\hline
\end{tabular}
\vspace{-0.4cm}
\end{table}

\section{Conclusion}


We presented a systematic analysis of discrete unit based vocoders in a multilingual multi-speaker setting by extending BigVGAN to synthesize speech from discrete units. Experiments across four Indian languages show that cluster size primarily controls intelligibility through phonetic resolution, while explicit speaker conditioning is essential for preserving speaker identity. At smaller cluster sizes, acoustically similar phonemes across languages share the same cluster IDs, while larger clusters separate them into distinct units. Consequently, language supervision provides additional gains at smaller cluster sizes where unit representations remain coarse. Future work will extend this analysis to all 22 Indian languages and explore richer conditioning strategies such as pitch, prosody, and other controllable speech attributes.

\section{Generative AI Use Disclosure}
Generative AI was utilized exclusively for minor linguistic refinement and phrasing improvements. The authors developed and verified all scientific concepts, experimental methodologies, results, and conclusions independently.
\bibliographystyle{IEEEtran}
\bibliography{mybib}

\end{document}